\documentclass{Interspeech}

\interspeechcameraready 

\usepackage{cite}
\usepackage{amsmath,amssymb,amsfonts}
\usepackage{graphicx} 
\usepackage{textcomp}
\usepackage{booktabs} 
\usepackage{stfloats} 

\usepackage{xurl}
\hypersetup{breaklinks=true}
\title{IMSE: Efficient U-Net-based Speech Enhancement using Inception Depthwise Convolution and Amplitude-Aware Linear Attention\thanks{This work was supported by the Research on Key Technologies of Data Security for Urban Gas Industrial IoT Systems Based on National Commercial Cryptography (Grant No. KJZD20230923114906013).}}

\author[affiliation={1}]{Xinxin}{Tang}

\author[affiliation={2}]{Bin}{Qin$^*$}

\author[affiliation={1}]{Yufang}{Li}

\affiliation{}{College of Electronics and Information Engineering, Shenzhen University}{Shenzhen, China}

\affiliation{}{Information Center, Shenzhen University}{Shenzhen, China}

\email{2510043011@mails.szu.edu.cn, qinbin@szu.edu.cn, 2510043056@mails.szu.edu.cn}


\newcommand\blfootnote[1]{%
  \begingroup
  \renewcommand\thefootnote{}\footnote{#1}%
  \addtocounter{footnote}{-1}%
  \endgroup
}

\begin{document}

\maketitle

\blfootnote{* Corresponding author.}

\begin{abstract}
Achieving a balance between lightweight design and high performance remains a significant challenge for speech enhancement (SE) tasks on resource-constrained devices. Existing state-of-the-art methods, such as MUSE, have established a strong baseline with only 0.51M parameters by introducing a Multi-path Enhanced Taylor (MET) transformer and Deformable Embedding (DE). However, MUSE suffers from structural redundancy due to its approximate-then-compensate attention mechanism and computationally expensive deformable offsets. We propose IMSE to resolve these bottlenecks via: 1) Amplitude-Aware Linear Attention (MALA), which fundamentally rectifies magnitude loss in linear attention, eliminating the need for auxiliary compensation branches; and 2) Inception Depthwise Convolution (IDConv), which efficiently captures the anisotropic features of spectrograms (e.g., harmonic strips) using decomposed static kernels instead of heavy dynamic deformations. Extensive experiments on the VoiceBank+DEMAND dataset demonstrate that IMSE significantly reduces the parameter count by 16.8\% (from 0.513M to 0.427M) while achieving superior performance compared to the MUSE baseline. Specifically, it achieves a PESQ score of \textbf{3.399}, setting a new state-of-the-art benchmark for ultra-lightweight speech enhancement models.
\end{abstract}

\section{Introduction}
\label{sec:introduction}

Speech enhancement (SE) algorithms, a core technology in the speech signal processing field, aim to recover clean speech from noise-corrupted signals, thereby improving speech quality and intelligibility. This technology has wide-ranging applications in mobile communications, hearing aid design, and automatic speech recognition (ASR) front-ends. In recent years, with the rapid development of deep learning, deep neural network (DNN)-based SE methods \cite{bulut2020low, tran2020single, botinhao2016investigating, moliner2022two} have achieved performance significantly superior to traditional signal processing methods in suppressing non-stationary noise and reverberation.

Currently, mainstream SE models typically adopt Encoder-Decoder or Two-Stage architectures, utilizing time-frequency (T-F) domain masking or complex spectral mapping for enhancement. To capture long-range contextual dependencies, the Transformer \cite{neil2020transformers, vaswani2017attention} and its variants have been widely introduced into SE tasks. However, the standard self-attention mechanism has a computational complexity of $O(N^2)$ (quadratic to the sequence length $N$), which is prohibitive for processing long speech sequences at high sampling rates. Furthermore, to achieve SOTA performance, existing models often stack a large number of layers and channels, leading to a surge in model parameters and computation, making real-time deployment on edge devices difficult.

To address these issues, Zizhen Lin et al. recently proposed MUSE \cite{lin2024muse}, a lightweight U-Net-based model (0.51M parameters). MUSE's success is attributed to two key designs: first, the use of Deformable Embedding (DE) to adapt to the irregular shapes of spectrogram patterns; second, the proposal of the Multi-path Enhanced Taylor (MET) transformer, which uses Taylor expansion to reduce attention complexity to linear and compensates for information loss with an additional convolutional branch.

Although MUSE achieves an impressive balance, we argue that its core components still have room for optimization:
\begin{enumerate}
    \item \textbf{Complexity of the MET module}: MET employs Taylor Self-Attention (T-MSA) \cite{qiu2023mb} as its core. However, Qihang Fan et al. \cite{fan2025rectifying} pointed out that standard linear attention formulas lose the amplitude information of the Query vector during normalization, leading to overly smooth and non-selective attention distributions. To compensate for this, MUSE must introduce a complex CSA branch, which adds structural redundancy.
    \item \textbf{Computational Overhead of the DE module}: Although the Deformable Star-shaped Dilated Convolution (DSDCN) \cite{qiu2023mb} in the DE module can flexibly adapt to spectrogram features, its offset generation and bilinear interpolation operations require additional computational resources during inference.
\end{enumerate}

In light of this, this paper proposes \textbf{IMSE}\footnote{Our open-source code is available at: \url{https://github.com/XinXinTang123/IMSE}}, aiming to further compress the model and maximize parameter efficiency by introducing more advanced algorithms. Our main contributions are as follows:
\begin{itemize}
    \item We introduce \textbf{Amplitude-Aware Linear Attention (MALA)} \cite{fan2025rectifying} to replace the MET module. MALA, through a novel normalization strategy, reintroduces the Query's amplitude information, enabling linear attention to produce sharp and accurate distributions similar to Softmax attention. This maintains strong global modeling capabilities even after removing the compensation branch.
    \item We employ \textbf{Inception Depthwise Convolution (IDConv)} \cite{yu2024inceptionnext} to replace the DSDCN in the DE module. IDConv decomposes large-kernel convolutions into multiple parallel small-kernel and strip convolutions, maintaining a large receptive field and enhancing anisotropic feature extraction while reducing parameters and computational complexity.
    \item Experimental results on the VoiceBank+DEMAND dataset show that IMSE, with only 0.427M parameters, achieves a SOTA trade-off between performance and efficiency, surpassing the original MUSE model in terms of parameter efficiency while maintaining superior speech quality.
\end{itemize}

\section{Related Work}
\label{sec:related_work}

\subsection{Efficient Speech Enhancement Networks}
Early deep SE methods relied mainly on CNNs and RNNs. To handle long sequences, TSTNN \cite{wang2021tstnn} proposed a two-stage Transformer architecture. To reduce computation, MP-SENet \cite{lu2023mp} adopted a strategy of processing magnitude and phase in parallel. MUSE \cite{lin2024muse} further combined the U-Net architecture with a linear Transformer, exploring an ultra-lightweight design under 1M parameters. Our work continues this line of research, focusing on discovering more efficient fundamental algorithms.

\subsection{Linear Attention Mechanisms}
The quadratic complexity of the standard Transformer is its main bottleneck in long-sequence tasks. Linear attention reduces the $O(N^2)$ complexity to $O(N)$ using the kernel trick. Common approaches include using $\phi(x) = \text{elu}(x)+1$ or ReLU as the kernel function. However, these methods often overlook the amplitude information loss caused by the non-linear activation. MALA, recently proposed by Qihang Fan et al. \cite{fan2025rectifying}, deeply analyzes this "amplitude-ignoring" phenomenon and proposes a magnitude-preserving computation paradigm, significantly boosting linear attention performance. IMSE leverages this latest advancement to streamline the network structure.

\subsection{Large-Kernel Convolution and Decomposition}
To expand the receptive field of CNNs, researchers have proposed large-kernel convolutions (e.g., $7\times7$ or larger). However, these have high parameter and computational costs. InceptionNeXt \cite{yu2024inceptionnext} proposed an Inception-based decomposition strategy, splitting a large-kernel depthwise convolution into multiple branches, such as $K\times K$, $1\times K$, and $K\times 1$. This design not only reduces computational cost but its strip convolution kernels are highly suitable for capturing features along the time axis (duration) and frequency axis (harmonic structures) in spectrograms, aligning well with the physical properties of speech signals.

\section{Proposed Method}
\label{sec:method}

\begin{figure*}[t]
    \centering
    \includegraphics[width=0.9\linewidth, height=6cm]{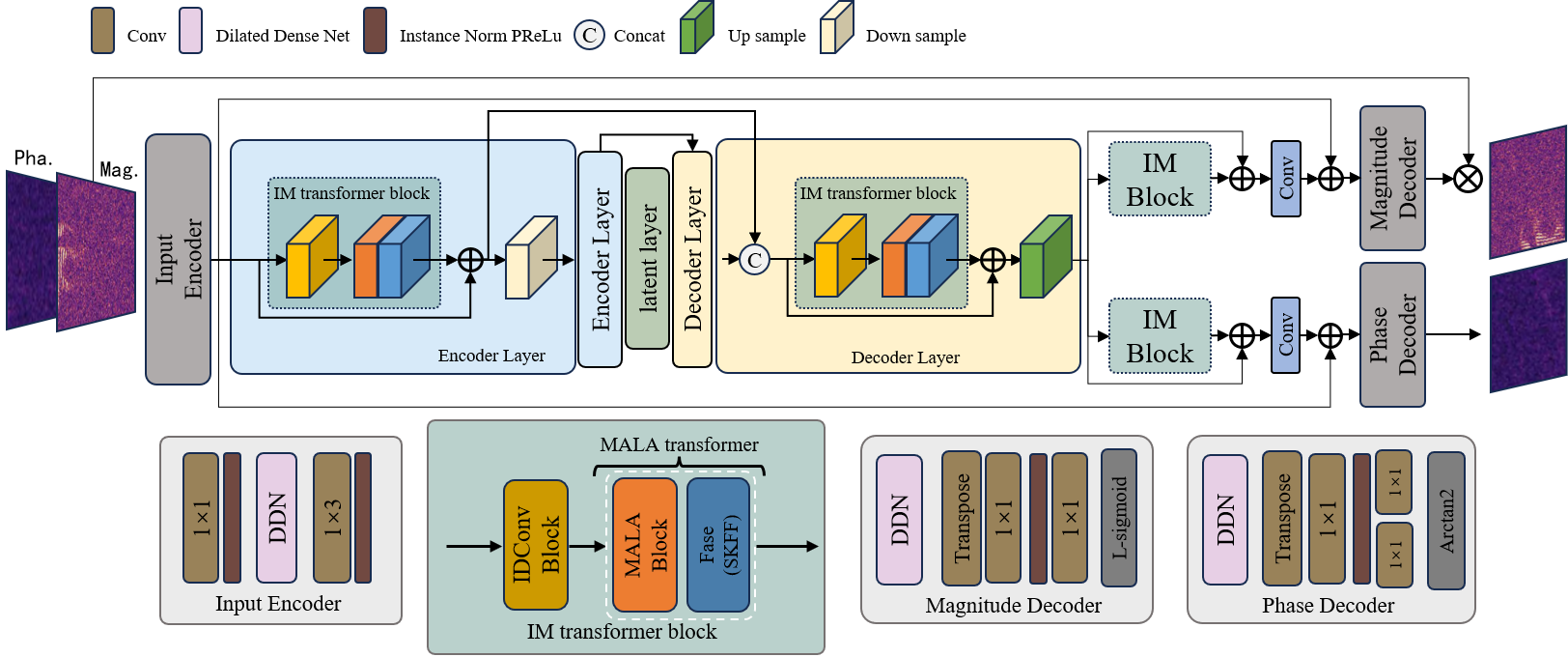}
    \caption{\textbf{The overall architecture of IMSE.}}
    \label{fig:architecture}
\end{figure*}

The overall architecture of IMSE is shown in Fig. \ref{fig:architecture}. It retains the four-level U-Net backbone of MUSE, processing complex-valued features from the STFT. This section focuses on the two core modules we replaced: Inception Depthwise Convolution Embedding (IDConv) and Amplitude-Aware Linear Attention (MALA).

\subsection{Amplitude-Aware Linear Attention (MALA)}
The MET module approximates Softmax via Taylor expansion but sacrifices the Query's magnitude information, resulting in over-smoothed attention. Consequently, MUSE requires a redundant branch to compensate. We instead adopt MALA, which mathematically rectifies this flaw by reintroducing magnitude via a division-based normalization. This restores the sharpness of Softmax attention at $O(N)$ complexity without structural redundancy.

To address this, we replace MET with Amplitude-Aware Linear Attention (MALA) \cite{fan2025rectifying}. The core idea of MALA is to reintroduce the amplitude information of the query $Q$ into linear attention, allowing it to mimic Softmax properties while maintaining $O(N)$ linear complexity.

MALA abandons traditional addition-based normalization (like Softmax) and instead adopts a division-based normalization method. It first maps $Q$ and $K$ through a non-linear kernel $\phi(\cdot)$ (e.g., $\phi(x) = \text{elu}(x) + 1$) to ensure non-negative attention scores.

For the $i$-th query $Q_i$, MALA introduces a scaling factor $\beta$ and an offset term $\gamma$. The attention score is defined as:
\begin{equation}
    \text{Attn}(Q_i, K_j) = \beta\phi(Q_i)\phi(K_j)^T - \gamma
\end{equation}
where $\beta$ and $\gamma$ are dynamically calculated based on the magnitude of $Q_i$, defined as:
\begin{align}
    \beta &= 1 + \frac{1}{\phi(Q_i) \sum_{m=1}^{N} \phi(K_m)^T} \\
    \gamma &= \frac{{\phi(Q_i) \sum_{m=1}^{N} \phi(K_m)^T}}{N}
\end{align}
Note that the denominator $\phi(Q_i) \sum_{m=1}^{N} \phi(K_m)^T$ is a scalar that captures the interaction between $Q_i$'s magnitude and all $K$.

Applying the attention scores to the values $V$, the complete computation of MALA is as follows:
\begin{align}
    Y_i &= \sum_{j=1}^{N} \text{Attn}(Q_i, K_j) V_j \\
    &= \sum_{j=1}^{N} (\beta\phi(Q_i)\phi(K_j)^T - \gamma) V_j \\
    &= \beta\phi(Q_i) \left( \sum_{j=1}^{N} \phi(K_j)^T V_j \right) - \gamma \left( \sum_{j=1}^{N} V_j \right)
\end{align}
Through this design, MALA successfully integrates the magnitude information of $Q_i$ into the calculation of $\beta$ and $\gamma$. As shown in the original paper, when the magnitude $\phi(Q_i)$ increases, $\beta$ and $\gamma$ adjust accordingly, making the attention distribution more concentrated and "sharp," consistent with the behavior of Softmax attention.

Finally, the computation in Eq. (6) utilizes the associative property of linear attention (i.e., $Q(K^TV)$), pre-calculating $\sum_{j=1}^{N} \phi(K_j)^T V_j$ and $\sum_{j=1}^{N} V_j$ as context vectors. This allows MALA to efficiently simulate the magnitude-aware properties of Softmax attention at $O(N)$ linear complexity, thereby precisely capturing global context in the speech signal.

\begin{figure}[th] 
    \centering
    \includegraphics[width=0.5\linewidth, height=5cm]{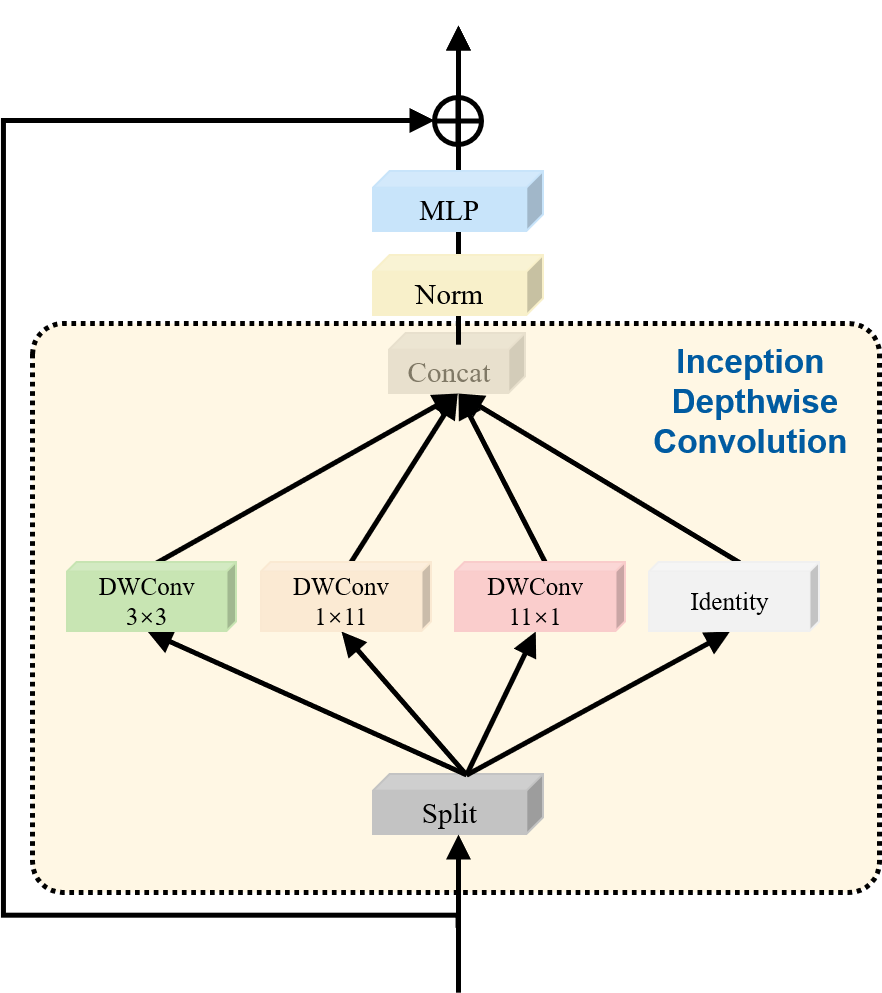}
    \caption{\textbf{Details of IDConv modules.} The figure shows how IDConv splits features for processing.}
    \label{fig:modules}
\end{figure}

\subsection{Inception Depthwise Convolution Embedding (IDConv)}
Speech spectrograms exhibit strong anisotropy—continuous temporal formants appear as horizontal strips, while wideband noise appears as vertical strips. Unlike MUSE's DSDCN which expends resources learning dynamic shapes, IDConv explicitly targets these physics-driven patterns using decomposed strip convolutions ($1\times K, K\times 1$). This captures acoustic textures with a static, lightweight structure, avoiding the latency of bilinear interpolation.

Inspired by InceptionNeXt \cite{yu2024inceptionnext}, we introduce IDConv as the feature embedding layer. IDConv splits the input channels into four groups, processed through different paths:
\begin{itemize}
    \item Branch 1 (Identity): An identity mapping to preserve original features and aid gradient propagation.
    \item Branch 2 (Square Kernel): Uses a $3\times3$ depthwise convolution (DW) to capture local time-frequency details.
    \item Branch 3 (Horizontal Band): Uses a $1\times 11$ depthwise (strip) convolution, specializing in capturing long-term dependencies along the time axis, such as speech duration features.
    \item Branch 4 (Vertical Band): Uses an $11\times 1$ depthwise (strip) convolution, specializing in capturing structures along the frequency axis, such as harmonic correlations.
\end{itemize}

Formally, let the input feature be $X$. We split it along the channel dimension into $X_1, X_2, X_3, X_4$. The output $Y$ is expressed as:
\begin{equation}
    Y = \text{Concat}(X_1, \text{DW}_{3\times3}(X_2), \text{DW}_{1\times11}(X_3), \text{DW}_{11\times1}(X_4))
\end{equation}
This decomposition strategy not only mimics the ability of deformable convolution to adapt to the elongated, "crescent-like" shapes of speech features in spectrograms \cite{lin2024muse}, but is also based entirely on static convolutions, greatly optimizing memory access and inference speed. Unlike square kernels used in visual tasks, speech spectrograms exhibit distinct anisotropic patterns (e.g., harmonic structures along the frequency axis and temporal continuity along the time axis). Our IDConv is specifically adapted to capture these acoustic characteristics using strip convolutions.

\section{Experiments}
\label{sec:experiments}

\subsection{Datasets and Experimental Setup}
We evaluate IMSE on the widely-used VoiceBank+DEMAND dataset \cite{botinhao2016investigating}.
\begin{itemize}
    \item \textbf{Dataset}: The training set contains 11,572 noisy utterances from 28 speakers, mixed with 10 noise types (at SNRs of 0, 5, 10, 15 dB). The test set contains 824 utterances from 2 unseen speakers (at SNRs of 2.5, 7.5, 12.5, 17.5 dB).
    \item \textbf{Preprocessing}: All audio is downsampled to 16kHz. STFT uses a 32ms frame length (510 points) and a 16ms frame shift (256 points).
    \item \textbf{Training Details}: We use the AdamW optimizer for 100 epochs with an initial learning rate of $5 \times 10^{-4}$. To ensure an ultra-lightweight model, we set the model's base channel count to 16.
\end{itemize}

\subsection{Evaluation Metrics}
We use five standard objective metrics to evaluate the quality of the enhanced speech:
\begin{itemize}
    \item \textbf{PESQ}: Perceptual Evaluation of Speech Quality (Range: -0.5 to 4.5).
    \item \textbf{CSIG}: Mean Opinion Score (MOS) prediction of signal distortion (Range: 1 to 5).
    \item \textbf{CBAK}: MOS prediction of background noise interference (Range: 1 to 5).
    \item \textbf{COVL}: MOS prediction of overall quality (Range: 1 to 5).
    \item \textbf{STOI}: Short-Time Objective Intelligibility (Range: 0 to 1).
\end{itemize}

\begin{table*}[htbp] 
    \caption{Quantitative comparison with other SOTA methods on the VoiceBank+DEMAND dataset. "-" indicates the metric was not reported in the original paper. Bold indicates the best result.}
    \label{tab:sota}
    \centering
    \small
    \setlength{\tabcolsep}{10pt} 
    \begin{tabular}{lccccccc}
        \toprule
        \textbf{Method} & \textbf{Year} & \textbf{Params (M)} & \textbf{PESQ} & \textbf{CSIG} & \textbf{CBAK} & \textbf{COVL} & \textbf{STOI} \\
        \midrule
        Noisy & - & - & 1.97 & 3.35 & 2.44 & 2.63 & 0.91 \\
        SEGAN \cite{pascual2017segan} & 2017 & 43.2 & 2.16 & 3.48 & 2.94 & 2.80 & 0.92 \\
        DEMUCS \cite{defossez2019demucs} & 2019 & 33.53 & 3.07 & 4.31 & 3.40 & 3.63 & 0.95 \\
        SE-Conformer \cite{kim2021se} & 2021 & - & 3.13 & 4.45 & 3.55 & 3.82 & 0.95 \\
        MetricGAN+ \cite{fu2021metricgan+} & 2021 & - & 3.15 & 4.14 & 3.16 & 3.64 & - \\
        TSTNN \cite{wang2021tstnn} & 2021 & 0.92 & 2.96 & 4.33 & 3.53 & 3.67 & 0.95 \\
        DB-AIAT \cite{yu2022dual} & 2022 & 2.81 & 3.31 & 4.61 & 3.75 & 3.96 & - \\
        DPT-FSNet \cite{dang2022dpt} & 2022 & 0.88 & 3.33 & 4.58 & 3.72 & 4.00 & \textbf{0.96} \\
        PHASEN \cite{yin2020phasen} & 2020 & 20.9 & 2.99 & 4.18 & 3.45 & 3.50 & 0.95 \\
        MetricGAN-OKDv2 \cite{shin2023metricgan} & 2023 & 0.82 & 3.12 & 4.27 & 3.16 & 3.71 & 0.95 \\
        MANNER \cite{park2022manner} & 2022 & 24.07 & 3.21 & 4.53 & 3.65 & 3.91 & 0.95 \\
        MANNER-S-5.3GF \cite{shin2022multi} & 2022 & 0.90 & 3.06 & 4.42 & 3.58 & 3.77 & 0.95 \\
        MUSE (Baseline) \cite{lin2024muse} & 2024 & 0.513 & 3.370 & 4.63 & 3.80 & 4.10 & 0.95 \\
        \midrule
        \textbf{IMSE (Ours)} & \textbf{2025} & \textbf{0.427} & \textbf{3.399} & \textbf{4.67} & \textbf{3.83} & \textbf{4.14} & {0.95} \\
        \bottomrule
    \end{tabular}
\end{table*}

\subsection{Comparison with State-of-the-Art Methods}
Table \ref{tab:sota} presents the comparison of IMSE with other advanced SE methods. We select classic models like SEGAN, TSTNN, and our baseline, MUSE, for comparison.

As shown in Table 1, IMSE achieves a PESQ score of \textbf{3.399} with only 0.427M parameters.
\begin{enumerate}
    \item \textbf{Comparison with baseline MUSE:} IMSE not only reduces the parameter count by 16.8\% (0.427M vs. 0.513M) but also improves the speech quality across all metrics. Notably, PESQ increases from 3.370 to 3.399, and CSIG/COVL see significant gains (4.67 vs 4.63, 4.14 vs 4.10). This demonstrates that our lightweight design extracts features more effectively than the redundant structures in MUSE.
    \item \textbf{Comparison with other lightweight models}: Compared to TSTNN (0.92M) and DPT-FSNet (0.88M), IMSE achieves significantly better performance (e.g., PESQ 3.373 vs. 2.96) with roughly half the parameters.
\end{enumerate}

\subsection{Ablation Study}
To verify the effectiveness of our two core components (MALA and IDConv), we conducted a detailed ablation study under the same experimental settings. The results are shown in Table \ref{tab:ablation}.

\begin{itemize}
    \item \textbf{Effectiveness of MALA}: When we replaced MET in MUSE with MALA (Row 2), the parameter count dropped significantly to 0.438M (a 14.6\% reduction from 0.513M). Surprisingly, performance did not degrade but slightly improved on PESQ (3.372). This confirms that MALA, by rectifying the 'amplitude-ignoring' issue, maintains strong modeling power while removing the redundant compensation branch.
    \item \textbf{Effectiveness of IDConv}: When we only replaced the DSDCN module in DE with IDConv (Row 3), parameters decreased slightly (0.501M), but PESQ improved significantly to 3.392. This suggests that multi-scale convolutional decomposition captures spectrogram texture features more effectively than deformable convolution alone.
    \item \textbf{Combined Effect:} IMSE (Row 4) integrates both MALA and IDConv. Remarkably, this combination achieves the highest performance across all metrics, with a PESQ of \textbf{3.399}, surpassing both the baseline and the single-module variants (e.g., 3.392 for IDConv-only). 

This result reveals a strong \textbf{synergistic effect}: MALA captures global amplitude-aware dependencies, while IDConv meticulously extracts anisotropic local spectral features. The combination allows the model to achieve peak performance with the lowest parameter count (0.427M), proving that efficiency and high-quality reconstruction can be achieved simultaneously without compromise.
\end{itemize}
\begin{table}[th] 
    \caption{Ablation study on component contributions. Validating the impact of MALA and IDConv on model parameters and performance.}
    \label{tab:ablation}
    \centering
    \scriptsize
    \setlength{\tabcolsep}{0.5pt} 
    \begin{tabular}{lccccc}
        \toprule
        \textbf{Model Configuration} & \textbf{Params (M)} & \textbf{PESQ} & \textbf{CSIG} & \textbf{CBAK} & \textbf{COVL} \\
        \midrule
        1. MUSE (Base) & 0.513 & 3.370 & 4.63 & 3.80 & 4.10 \\
        2. + MALA (replaces MET) & 0.438 & 3.372 & 4.61 & 3.80 & 4.09 \\
        3. + IDConv (replaces DE) & 0.501 & 3.392 & 4.65 & 3.82 & 4.13 \\
        4. \textbf{IMSE (Ours)} & \textbf{0.427} & \textbf{3.399} & \textbf{4.67} & \textbf{3.83} & \textbf{4.14} \\
        \bottomrule
    \end{tabular}
\end{table}

\section{Conclusion}
\label{sec:conclusion}

This paper proposed IMSE, an ultra-lightweight speech enhancement network. Targeting the efficiency bottlenecks in the existing SOTA model MUSE, we utilized Inception Depthwise Convolution (IDConv) to efficiently capture multi-scale spectrogram features and Amplitude-Aware Linear Attention (MALA) to resolve the flaws of traditional linear attention. Experimental results strongly demonstrate that IMSE achieves an optimal balance between lightweight design and high performance. It achieves a PESQ score of \textbf{3.399} with an extremely low parameter count of 0.427M, establishing a new benchmark for efficiency. Future work will focus on deploying this model on actual low-power chips and exploring its application in real-time speech communication.



\end{document}